\documentstyle[prl,aps,12pt]{revtex}

\begin{document}
\author{A. P\'{e}rez-Madrid}
\title{A SIMPLE MODEL FOR NONEXPONENTIAL RELAXATION IN COMPLEX DYNAMICS}
\address{Departament de F\'{\i}sica Fonamental. \\
Facultat de F\'{\i}sica. Universitat de Barcelona. \\
Diagonal 647, Barcelona. Spain}
\maketitle

\begin{abstract}
The nonexponential relaxation ocurring in complex dynamics manifested in a
wide variety of systems is analyzed through a simple model of diffusion in
phase space. It is found that the inability of the system to find its
equilibrium state in any time scale becomes apparent in an effective
temperature field which leads to a hierarchy of relaxation times responsible
for the slow relaxation phenomena.
\end{abstract}

\pacs{\# }

{PACS numbers: 64.70.Pf, 02.50.Ey, 05.10.Gg}

Complex dynamics is the object of active research due to its implications in
technology of materials and in several fields of scientific knowledge. At
the physicochemical and biological level, complex dynamics is observed in
glass-forming liquids, mechanical, dielectric and magnetic relaxation,
amorphous semiconductors, pinned density-wave, protein dynamics, protein
folding and population dynamics among others. The mechanisms underlying slow
relaxation in complex dynamics still lack a clear and definitive
elucidation. In the case of supercooled liquids and glasses, several
experiments and computer simulations have been done which support the
explanation of these relaxation phenomena in the framework of the energy
landscape paradigm as the result of activated diffusion through a rough
energy landscape of valleys and peaks \cite{stillinger}-\cite{ediger}.

To understand what these mechanisms are, we propose here a simple model to
show a possible origin of nonexponential relaxation based on the idea of the
energy landscape and nontrivial energy barriers. This model which consists
of the diffusion in phase space, provides a direct link between the phase
space dynamics and the slow relaxation of the functions of the configuration
of the system in the corresponding energy landscape. The slowing down of the
dynamics clearly appears as a consequence of the freezing of some degrees of
freedom which takes the system out of equilibrium. This fact is indicated by
the presence of an effective temperature field incorporating the information
of the suppressed degrees of freedom and depending on the equilibrium
temperature at the moment the quench was applied.

Hence, we model the relaxation in the liquid as the Brownian motion of a
test particle of unit mass in a potential. As it is well known, that
physical situation is \ described by the Klein-Kramers equation which for
simplicity we write in one dimension 
\begin{equation}
\frac{\partial }{\partial t}\widehat{\rho }=-\frac{\partial }{\partial x}u%
\widehat{\rho }+\frac{\partial }{\partial u}\widehat{\rho }\frac{\partial }{%
\partial x}V(x)+\gamma \frac{\partial }{\partial u}\left( \beta ^{-1}\frac{%
\partial }{\partial u}+u\right) \widehat{\rho }\text{ ,}  \label{kramer}
\end{equation}
where $\widehat{\rho }(x,u,t)$ is the probability density, $x$ and $u$ are
the position and velocity of the test particle, $\gamma $ its friction
coefficient and $\beta ^{-1}=k_{B}T_{o}$, with $k_{B}$ being the Boltzmann
constant and $T_{o}$ the bath temperature. Here, $V(x)$ is a nonperiodic
potential constituting a schematic representation of a rough energy
landscape. The local equilibrium solution for the Kramers problem (\ref%
{kramer}) is the Boltzmann distribution $\widehat{\rho }_{l.eq.}(x,u)\sim
\exp \left\{ -\beta \left( 1/2u^{2}+V(x)\right) \right\} $. Thus, inherent
to Eq.\ (\ref{kramer}) is the existence of local equilibrium in phase space $%
(x,u)$ \cite{rubi}, \cite{vilar}, and the approach to this state occurs at
the bath temperature. Therefore, one concludes that at high temperatures the
system possesses a large amount of energy to move through the whole phase
space.

We assume that a quench of the system freezes the traslational degrees of
freedom taking the system away from equilibrium. Thus, this leads us to
think of the partial decoupling of the probability density $\widehat{\rho }%
(x,u,t)$ 
\begin{equation}
\widehat{\rho }(x,u,t)=\phi _{x}(u,t;T_{o})\rho (x,t)\text{ ,}
\label{factorization}
\end{equation}
where the conditional probability $\phi _{x}(u,t;T_{o})$ describes a state
of quasi-equilibrium \cite{franz}, \cite{ivan} in which the system, unable
to equilibrate at the bath temperature, remains hanging. Assuming that $u$
is the fast variable, one concludes that the dynamical processes in the
system are associated to configurational changes related to $x$ which
constitutes the slow variable whose probability density is $\rho (x,t)=\int 
\widehat{\rho }du$. Hence, by integration of Eq. (\ref{kramer}) the time
derivative of $\rho $ is obtained 
\begin{equation}
\frac{\partial }{\partial t}\rho =-\frac{\partial }{\partial x}\int u%
\widehat{\rho }du\text{ }  \label{continuity}
\end{equation}
which defines the current $J(x,t)=\int u\widehat{\rho }du.$ This current
evolves according to 
\begin{equation}
\frac{\partial }{\partial t}J=-\frac{\partial }{\partial x}\int u^{2}%
\widehat{\rho }du\text{ }+\frac{\partial }{\partial x}V\int u\frac{\partial 
}{\partial u}\widehat{\rho }du+\gamma \int u\frac{\partial }{\partial u}%
\left( \beta ^{-1}\frac{\partial }{\partial u}+u\right) \widehat{\rho }du
\label{derivadaj}
\end{equation}
which has been obtained from Eq. (\ref{kramer}). Then, after partial
integration of Eq. (\ref{derivadaj}) and using the decoupling approximation (%
\ref{factorization}), for times $t\gg \gamma ^{-1}$ we obtain 
\begin{equation}
J(x,t)=-\tau \left\{ \rho (x,t)\frac{\partial }{\partial x}V(x)+k_{B}\frac{%
\partial }{\partial x}\rho (x,t)T(x,t;T_{o})\right\} \text{ ,}
\label{current}
\end{equation}
where $\tau =\gamma ^{-1}$ and $T(x,t;T_{o})=\int u^{2}\phi
_{x}(u,t;T_{o})du $ is the second moment of the conditional distribution $%
\phi _{x}(u,t;T_{o})$ which plays the role of an effective temperature which
contains information on the frozen degrees of freedom. After defining the
effective potential $\Phi (x,t)=V(x)+k_{B}T(x,t;T_{o})$ Eq. (\ref{current})
can be rewritten as 
\begin{equation}
J(x,t)=-D(x,t;T_{o})\frac{\partial }{\partial x}\rho (x,t)-\tau \rho (x,t)%
\frac{\partial }{\partial x}\Phi (x,t)\text{ ,}  \label{current2}
\end{equation}
where $D(x,t;T_{o})=\tau k_{B}T(x,t;T_{o})$ is the generalized diffusion
coefficient. Thus, by substituting Eq. (\ref{current2}) into Eq. (\ref%
{continuity}) and taking $\tau =1$ ({\em i.e. }rescaling the time t), this
becomes the generalized diffusion equation 
\begin{equation}
\frac{\partial }{\partial t}\rho =\frac{\partial }{\partial x}\left\{
D(x,t;T_{o})\frac{\partial }{\partial x}\rho (x,t)+\rho (x,t)\frac{\partial 
}{\partial x}\Phi (x,t)\right\} \text{ ,}  \label{diffusionequation}
\end{equation}
Note that the temperature field $T(x,t;T_{o})$ introduces thermal barriers
(or in others words non trivial activation energies) in the system \cite%
{sastry}. The presence of these thermal barriers is a nonequilibrium effect
that disappears when the system is in equilibrium. In addition, as a
consequence of the elimination of degrees of \ freedom, the dynamics becomes
non-Markovian and depends on the equilibrium temperature $T_{o}$ at the time
of quench. By introducing the Massieu function 
\begin{equation}
\Theta (x,t)=\log \rho (x,t)+\int \frac{1}{k_{B}T(x,t;T_{o})}\frac{\partial 
}{\partial x}\Phi (x,t)dx  \label{massieu}
\end{equation}
the probability current Eq. (\ref{current2}) can be rewritten as 
\begin{equation}
J(x,t)=-D(x,t;T_{o})\rho _{l.qe.}(x,t)\frac{\partial }{\partial x}\exp
\Theta (x,t)\text{ ,}  \label{current3}
\end{equation}
where $\rho _{l.qe.}(x,t)\sim \exp \left\{ -\int \frac{1}{k_{B}T(x,t)}\frac{%
\partial }{\partial x}\Phi (x,t)dx\right\} $ is the local quasi-equilibrium
probability density for which $J(x,t)=0$ or equivalently $\Theta (x,t)=$
constant.

If in a point $x_{2}$ there is a sink where $\rho (x_{2},t)=0$ and a source
at $x_{1}$ so that a quasi-stationary current $J(t)$ can be established in
the system, thus, integration of Eq. (\ref{current3}) gives us the
quasi-stationary probability density 
\begin{equation}
\rho (x,t)=J(t)\rho _{l.qe.}(x,t)\int_{x_{2}}^{x}\frac{dy}{D(y,t;T_{o})\rho
_{l.qe.}(y,t)}\text{ .}  \label{quasistationarydensity}
\end{equation}
On the other hand, a second integration leads to 
\begin{equation}
J(t)=n(t)\left( \int_{x_{1}}^{x_{b}}dx\rho _{l.qe.}(x,t)\int_{x}^{x_{2}}%
\frac{dy}{D(y,t;T_{o})\rho _{l.qe.}(y,t)}\right) ^{-1}\text{ ,}
\label{quasistationarycurrent}
\end{equation}
with $n(t)=\int_{x_{1}}^{x_{b}}\rho (x,t)dx$ being the population at the
left of the sink. Thus, we can formulate the following rate equation 
\begin{equation}
\frac{d}{dt}n(t)=-K(t;x_{b})n(t)\text{ ,}  \label{rateequation}
\end{equation}
where 
\begin{equation}
K(t;x_{b})=\left( \int_{x_{1}}^{x_{b}}dx\rho _{l.qe.}(x,t)\int_{x}^{x_{2}}%
\frac{dy}{D(y,t;T_{o})\rho _{l.qe.}(y,t)}\right) ^{-1}  \label{rateconstant}
\end{equation}
is the rate constant, and $x_{b}$ is the position of the barrier between $%
x_{1}$ and $x_{2}$. The relaxation equation (\ref{rateequation}) admits the
solution 
\begin{equation}
n(t)=n(t_{o})\exp \left\{ -\int_{t_{o}}^{t}K(t^{\prime };x_{b})dt^{\prime
}\right\} \equiv n(t_{o})\text{ }\exp \left\{ -f(t)\right\}
\label{population}
\end{equation}
or 
\begin{equation}
\widehat{n}(t)=\widehat{n}(t_{o})\text{ }\exp \left\{ -f(t/\tau )\right\} 
\text{ }  \label{population2}
\end{equation}
which constitutes our main result, where $\widehat{n}(t)=n(t/\tau )$ and the
initial time $t_{o}$ is known in the literature as the waiting time. Here,
the function $f(t/\tau )$ might be interpreted as an algebraic function $%
(t/\tau )^{\beta }$ obtaining a stretched exponential behavior \cite{angell}
or as a logarithmic function $-\alpha \log At$ which leads to a power law
behavior \cite{hu} that characterizes anomalous diffusion. In addition,
another reading of our result is possible since assuming the form of $%
f(t/\tau )$ is equivalent to assuming the form of the distribution of
residence times $\psi (t)\equiv -\frac{d}{dt}n(t)$ \cite{gezelter}. Thus, an
in-depth analysis of the implications of the supposition of a nonexponential
distribution of residence times usually performed in Continuous Time Random
Walk models, reveals that this hypothesis might also be rooted in the
nonequilibrium character of the dynamics inherent to the energy landscape
picture as we have shown.

Our previous discussion applies to fragile glasses. In the case of strong
glassy systems, the potential is almost periodic. Thus, when the temperature
is lowered, any configurational change is preceded by an equilibration in
momentum space given by a local Maxwellian. Hence, the activation energies
are independent from the temperature but with a certain random spatial
distribution. Averaging over this distribution of energy barrier gives one
the corresponding relaxation \cite{ediger}, \cite{berne}, \cite{wolynes2}.

We conclude that despite its simplicity, the model we study contains the
main features observed in the dynamical slowing down observed in a wide
variety of complex systems. Hence, we think this work will contribute to the
understanding of some aspects complex dynamics.

This work was supported by DGICYT of the Spanish Government under Grant No.
PB2002-01267.

\end{document}